\documentclass[12pt]{article}
\usepackage{a4wide}
\usepackage{amssymb}
\begin{document}
{\renewcommand{\thefootnote}{\fnsymbol{footnote}}
\hfill  IGPG--06/4--2\\
\medskip
\hfill gr--qc/0604105\\
\medskip
\begin{center}
{\LARGE  Singularities in Isotropic\\ Non-Minimal Scalar Field Models}\\
\vspace{1.5em}
Martin Bojowald\footnote{e-mail address: {\tt bojowald@gravity.psu.edu}}
and Mikhail Kagan\footnote{e-mail address: {\tt mak411@psu.edu}}
\\
\vspace{0.5em}
Institute for Gravitational Physics and Geometry,\\
The Pennsylvania State
University,\\
104 Davey Lab, University Park, PA 16802, USA\\
\vspace{1.5em}
\end{center}
}

\setcounter{footnote}{0}

\newcommand{\md}{{\mathrm d}}
\newcommand{\lp}{\ell_{\mathrm P}}
\newcommand{\be}{\begin{equation}}
\newcommand{\ee}{\end{equation}}
\newcommand{\bq}{\begin{eqnarray}}
\newcommand{\eq}{\end{eqnarray}}

\newcommand{\rcl}{\rho_{\mathrm{cl}}}
\newcommand{\rqm}{\rho_{\mathrm{q}}}
\newcommand{\rsc}{\rho_{\mathrm{sc}}}
\newcommand{\wcl}{w_{\mathrm{ cl}}}
\newcommand{\weff}{w_{\mathrm{eff}}}
\newcommand{\wsc}{w_{\mathrm{sc}}}
\newcommand{\wqm}{w_{\mathrm{q}}}
\newcommand{\mpl}{M_{\rm P}}
\newcommand{\lra}{\longrightarrow}
\newcommand{\ip}[2]{{\langle#1\,|\,#2\rangle}}

\def\bra#1{\mathinner{\langle{#1}|}}
\def\ket#1{\mathinner{|{#1}\rangle}}
\def\braket#1{\mathinner{\langle{#1}\rangle}}
\def\Bra#1{\left<#1\right|}
\def\Ket#1{\left|#1\right>}
\def\q{{}^o\!q}
\def\H{{\cal H}}
\def\L{\mu}
\def\f{\frac}
\def\t{\tilde}

\begin{abstract}
 Non-minimally coupling a scalar field to gravity introduces an
 additional curvature term into the action which can change the
 general behavior in strong curvature regimes, in particular close to
 classical singularities. While one can conformally transform any
 non-minimal model to a minimally coupled one, that transformation can
 itself become singular. It is thus not guaranteed that all
 qualitative properties are shared by minimal and non-minimal
 models. This paper addresses the classical singularity issue in
 isotropic models and extends singularity removal in quantum gravity
 to non-minimal models.
\end{abstract}

\section{Introduction}

It is well-established that general relativity implies space-time
singularities for most solutions deemed to be relevant for physical
situations such as cosmology or black holes
\cite{SingTheo,HawkingEllis}. Precise general properties of such
singularities, besides the criterion of geodesic incompleteness
usually used to prove singularity theorems, are however difficult to
find. Even in isotropic models there are several types of
singularities in addition to the common big bang type singularities at
vanishing scale factor $a(t)$ of a Friedmann--Robertson--Walker model
\begin{equation}\label{FRW}
 \md s^2=-\md t^2+a(t)^2\left(\frac{\md r^2}{1-kr^2}+
r^2\md\Omega^2\right)\,.
\end{equation}
Other possibilities include big rip singularities where the scale
factor diverges in finite proper time, such as in superinflationary
solutions of constant equation of state parameter $w<-1$, or sudden
singularities at non-zero, finite $a$. These alternatives are rather
special: While a big bang singularity can be shown to arise for
isotropic models using only energy conditions and non-zero expansion
at one time, singularities at non-zero $a$ require additional
conditions to be satisfied by the matter ingredients. Such
singularities can therefore be obtained in phenomenological matter
models, but not so easily in models sourced by scalar or other
fields. In minimally coupled scalar field models, for instance, one
has only the freedom to choose a potential which does not allow, e.g.,
singularities associated with super-inflationary expansion.

One additional freedom for scalar fields is a non-minimal coupling to
space-time curvature by adding a term $-\frac{1}{2}\xi R\phi^2$ to the
Lagrangian. This can change in particular the behavior close to
curvature singularities where $R$ becomes large. Such a term is
also interesting because it is the only possible curvature coupling of
any matter field which does not require one to use additional length
scales for the correct dimensions. Often, curvature couplings of
fundamental fields are considered as effective terms to be derived
from a quantum theory of gravity. In such a case, there is an
additional length parameter provided by the Planck length $\ell_{\rm
P}=\sqrt{G\hbar}$ which multiplies effective curvature terms. All
these terms thus vanish in the classical limit except for the
non-minimal coupling term of a scalar with quadratic coupling
function, which is not multiplied by $\ell_{\rm P}$. Since a classical
theory with such a term makes sense, one can also ask whether it would
change any results of quantum gravity obtained with minimal
coupling. Other curvature couplings, which are multiplied by powers of
$\ell_{\rm P}$, cannot reasonably be quantized because they arise only
from an effective description of an already quantized theory of
gravity.

One application of quantum gravity is singularity removal which can
depend on the matter type. Moreover, with a non-minimal coupling term
of a scalar already the classical behavior can change close to
curvature singularities. We therefore perform in this paper an
analysis of isotropic singularities in non-minimal scalar field models
based on canonical transformations to a minimally coupled model. We
will end with conclusions for singularity removal by quantum gravity.

\section{Non-minimal coupling}

Non-minimal coupling is usually introduced at the Lagrangian level
which makes properties of Lorentz invariance manifest. At the level of
the action one can also see that any non-minimally coupled model
allows a conformal transformation and re-definition of the field
variables which brings it into minimally coupled form
\cite{NonMinConf,Conformal}. This transformation, however, can be difficult to
determine analytically which makes the discussion of singularities
complicated. After recalling these properties, we will thus switch to
a canonical formulation. This allows us to perform canonical
transformations which are more general than transformations allowed at
the Lagrangian level. It will turn out that a
much simpler transformation to minimally coupled form is then possible
which we will exploit in our discussion of singularities.

\subsection{Equations of motion}

The simplest form of a non-minimally coupled scalar field action is
given by
\be \label{Action} S[g_{ab}, \phi]=\int{\md^4 x \sqrt{-g} \left(
\frac{f(\phi)}{2 \kappa} R - \frac{1}{2}g^{ab}\nabla_a \phi
\nabla_b \phi - U(\phi) \right)}
\ee
where $g \equiv \det(g_{ab})$, $\kappa = 8 \pi G$, $R$ is the
scalar curvature of the metric and $U(\phi)$ is the
self-interacting potential of the matter scalar field $\phi$. The
curvature term is coupled to matter through the coupling function
$f(\phi)$ in a form which is manifestly Lorentz invariant.
One can view this function as defining an
effective gravitational constant $\kappa$, in the sense
$\kappa_{\rm eff} = \kappa/f(\phi)$
and hence we require $f(\phi)$ to be always positive.
If $f$ is thought of as arising from a perturbative expansion of a
field theory, it is a polynomial and restricted by boundedness from
below to be an even function. Finally, the coupling function should
satisfy the limit $f(\phi) \to 1$ when $\phi
\to 0$, recovering the standard minimally coupled action.

Given the restrictions above, the simplest non-trivial form of the
coupling function is
\be \label{CplFcn}f(\phi)=1-\sigma \phi^2.\ee
In fact, this is the form of coupling function that is normally
considered in the literature. We will refer to $\sigma=\kappa\xi$ as
the coupling strength, vanishing $\sigma$ corresponding to the case of
minimal coupling. We will later see, however, that the behavior around
singularities can sensitively depend on the form of $f(\phi)$.

Upon reducing the action to a flat Friedmann--Robertson--Walker model
(\ref{FRW}) it becomes a functional only of the scale factor $a$ and
the homogeneous scalar $\phi$. Its canonical version depends on these
two values and their momenta. In what follows, we will use independent
variables $p\in{\mathbb R}$, such that $|p|=a^2$, and $c=\gamma\dot{a}$.
They have geometrical meaning as a triad component $p$ (such that
${\rm sgn}(p)$ determines the orientation of the triad) and connection
component $c=\gamma\dot{a}$ (or extrinsic curvature), which
corresponds to using isotropic Ashtekar instead of metric variables
\cite{AshVar,AshVarReell,IsoCosmo}. The positive real number $\gamma$
is the so-called Barbero--Immirzi parameter \cite{AshVarReell,Immirzi}
which we keep for generality but can be assumed to equal one
here. These variables are the basis for a loop quantization to be
employed in the final section.

With these variables the action (\ref{Action}) becomes
\be \label{action} S_{\rm sym}=\int{\md t \left[ \frac {3 f(\phi)}{\kappa
\gamma}  p \dot c + \pi \dot \phi - N \left(-\frac {3f(\phi)\sqrt{|p|}
c^2}{\gamma^2} + \f{\pi^2}{2 |p|^{3/2}} + |p|^{3/2} U(\phi)
\right)\right]},
\ee
with the lapse function $N$. The dot denotes a proper time
derivative. From (\ref{action}) we see that the pairs
$(c,f(\phi)p)$ and $(\phi,\pi \equiv |p|^{3/2} N^{-1} \dot \phi)$
are canonically conjugate variables with non-vanishing Poisson
brackets
\be \label{PB} \{c,f(\phi) p\}=\frac{\kappa \gamma}{3}, \quad
\{\phi, \pi\}=1\ee
Note that the Poisson bracket $\{p,\pi\}=f(\phi)p\{f(\phi)^{-1},\pi\}$
is not zero, in contrast to minimal coupling, while $\{f(\phi),p\}$
is. This will motivate our choice of the new canonical variables
transforming to a standard minimally coupled form later on. The
dynamics is generated by the Hamiltonian $H=N(H_G + H_\phi)$ where
\be H_G=-\frac{3 f(\phi)}{\kappa \gamma^2} \sqrt{|p|} c^2, \quad
H_\phi = \frac{\pi^2}{2 |p|^{3/2}}+|p|^{3/2} U(\phi)\,. \ee
It is now very convenient to proceed to conformal time by setting
the lapse $N:=\sqrt{|p|}$. The Hamiltonian constraint then becomes
\be \label{Ham} H=-\frac{3 f(\phi)}{\kappa \gamma^2} |p|
c^2+\frac{\pi^2}{2 |p|}+p^2 U(\phi)\,. \ee
\subsection{Conformal transformation}

A conformal transformation $\tilde{g}_{ab}=f(\phi)g_{ab}$ transforms
the action with non-minimal coupling to
\[
 S[\tilde{g}_{ab}, \phi]=\int{\md^4 x \sqrt{-\tilde{g}} \left(
\frac{1}{2 \kappa} \tilde{R} - \frac{1}{2}F(\phi)^2 \tilde{g}^{ab}
\tilde{\nabla}_a \phi
\tilde{\nabla}_b \phi - \tilde{V}(\phi) \right)}
\]
with
\[
 F(\phi)^2=\frac{1-(1-6\xi)\sigma\phi^2}{(1-\sigma\phi^2)^2}
\quad\mbox{ and }\quad
 \tilde{V}(\phi) = \frac{U(\phi)}{(1-\sigma\phi^2)^2}
\]
for a coupling function of the form (\ref{CplFcn}). This can be
brought to the standard minimally coupled form for the redefined
scalar $\tilde{\phi}=\int\md\phi F(\phi)$, implying a rather
complicated relation $\tilde{\phi}(\phi)$, e.g.,
\[
 \tilde{\phi}= \frac{1}{\kappa}\left(\sqrt{\xi^{-1}(1-6\xi)}
\arcsin\left(\sqrt{\sigma(1-6\xi)}\phi\right)+
\sqrt{\frac{3}{2}}\log\left|\frac{\sqrt{6\xi\sigma}\phi+
\sqrt{1-\sigma(1-6\xi)\phi^2}}{\sqrt{6\xi\sigma}\phi-
\sqrt{1-\sigma(1-6\xi)\phi^2}} \right|\right)
\]
if $\xi<1/6$.

From the canonical point of view, moreover, a conformal transformation
implies a more complicated change in the momenta conjugate to the
spatial metric. As we will see now, however, the larger freedom in
canonical transformations compared to conformal ones allows a similar
transformation to minimally coupled form which even leads to a simpler
redefinition of the scalar. This will be exploited in the rest of this
paper to discuss the behavior of isotropic singularities which will be
much more transparent than discussions based on conformal
transformations of the space-time metric.

\subsection{Canonical transformation}

In order to perform a canonical transformation
bringing the Hamiltonian into its standard (i.e.\ minimally coupled)
form, we start with the geometrical variables $(c,p)$ and define
\be \label{pRel}\t c := c, \quad \t p: = f(\phi) p
\ee
such that $\{\t c,\t p\}=\frac{\kappa \gamma}{3}$.  Note that this
does not correspond to a conformal transformation from the original
gravitational pair $(p,c)$ to a new one $(\tilde{p},\tilde{c})$
because in that case $c$ would have to change, too.  The Hamiltonian
thus takes the form
\be \label{HamCanon}
H=-\frac{3}{\kappa \gamma^2} |\t p| \t c^2+\frac{\pi^2
f(\phi)}{2 |\t p|}+\t p^2 \f {U(\phi)}{f(\phi)^2}. \ee
In this form, the Hamiltonian is written in terms of canonical
variables $(\tilde{p},\tilde{c},\phi,\pi)$ and would thus be the
starting point for a canonical quantization. Finally, redefining the
field variables
\be \label{phiDef} \t \pi := \sqrt{f(\phi)} \pi, \quad \t \phi :=
\int_0^\phi \! \f {d \phi^\prime}{\sqrt{f(\phi^\prime)}} \ee such
that $\{\t \phi,\t \pi \}=1$ we obtain
\be \label{newHam} H=-\frac{3}{\kappa \gamma^2} |\t p| \t
c^2+\frac{\t \pi^2}{2 |\t p|}+\t p^2 \t V(\t\phi) \ee where
\be \label{EffPot} \t V(\tilde{\phi}) :=
\f {U(\phi(\tilde{\phi}))}{f(\phi(\tilde{\phi}))^2}\ee
and $\phi$ is understood as a function of $\t\phi$ by inverting
(\ref{phiDef}). Specifically for the coupling function
(\ref{CplFcn}), the original and canonically transformed field are
related in a simple manner by
\begin{equation} \label{PhiRel}
\sqrt{\sigma} \phi = \sin(\sqrt{\sigma} \t\phi) \quad \mbox{for }
\sigma>0 \quad
{\rm and} \quad \sqrt{|\sigma|} \phi = \sinh({\sqrt{|\sigma|}
\t\phi}) \quad \mbox{for }\sigma<0\,.
\end{equation}
%
%
%
%

\subsection{Transformation singularities}

Both the conformal and canonical transformation can fail to be
one-to-one and thus become singular. For the conformal transformation
this happens when $\md\tilde\phi/\md\phi = F(\phi)=0$, i.e.\ at
\[
 \Phi_{\rm c}=\f{1}{\sqrt{\sigma(1-6\xi)}} \quad\mbox{for}\quad
0<\xi<\frac{1}{6}
\]
and when $F(\phi)$ diverges at $\phi_{\rm c}=1/\sqrt{\sigma}$ for $\xi>0$.
While $\Phi_{\rm c}$ turns out to be a spurious singularity
introduced by the transformation, $\phi_{\rm c}$ corresponds to a
physical shear singularity at least when anisotropic perturbations
are allowed \cite{NonMinAniso}.

The latter singularity clearly occurs for the canonical
transformation, too, since now
$\md\tilde{\phi}/\md\phi=f(\phi)^{-1/2}$ diverges at
$\phi_{\rm c}$. However, this is the only point where the transformation
fails to be one-to-one and there is no analog of $\Phi_{\rm c}$. The
canonical transformation thus avoids the unphysical regime
$\phi>\Phi_{\rm c}$ of the conformal transformation.

\section{Singularities}

We can now investigate what this implies for singularities since we
know the minimally coupled situation which is realized for the
transformed variables. Then, for a regular potential the singularity
corresponds to $\t p=0$ and $\t \phi \rightarrow \infty$
\cite{ScalarSing}. For $\sigma>0$, however, the potential can be
singular, thus changing properties of space-time singularities
\cite{ScalarSingWall}. In the original variables, the behavior of the
system can then be dramatically different for $\sigma<0$ and
$\sigma>0$ as is evident from properties of the transformation. We will
consider the former case first.

\subsection{Negative coupling constant}

For $\sigma<0$, the potential $\tilde{V}(\tilde{\phi})$ is regular
everywhere such that in the transformed variables singularities occur
in the usual form, $\tilde{p}=0$ and $\tilde{\phi}\to\infty$. For
purposes of quantization it will later be important to know what this
implies for the original canonical variables $\tilde{p}$ and $\phi$
since these are the variables that will be quantized. It follows from
the preceding considerations that the behavior of $\phi$ will not be
different from that of $\tilde{\phi}$ since the transformation
(\ref{PhiRel}) is one-to-one. In fact, in the vicinity of the
singularity we have $\phi \propto \exp(\sqrt{\sigma}\t \phi)$,
i.e. the original field also blows up. Interestingly, had the coupling
function been of the form $f(\phi)=1-\sigma \phi^{2+\epsilon}$
for some positive $\epsilon$, however small, the integral in
(\ref{phiDef}) would have been converging for $\phi \rightarrow
\infty$, that is a finite value of $\t\phi$ would have
corresponded to a diverging $\phi$ and the possibility of a
different type of singularity in the canonical variables would not
have been ruled out. The case of a quadratic coupling function,
which is preferred by physical arguments since it does not require
additional dimensionful constants, thus appears as the limiting
case where singularities in non-minimal models occur as in minimal
ones.

It will also be of interest to see the behavior in the geometrical
variables $p$ in addition to $\phi$.  Using the relation between the
scale factors (\ref{pRel}), we conclude that $\t p
\rightarrow 0$ implies $p \rightarrow 0$ since $f(\phi)\not=0$.
Thus, a singularity in the canonical variables implies a singularity
in the geometrical variables occurring at the same values of the field
and scale factor. The reverse conclusion is not true in
general since $p=0$ together with $\phi\to\infty$ could result in a
finite value of $\tilde{p}$. However, since we already know that the
canonical variables are regular except at $\tilde{p}=0$, such cases
are ruled out for $\sigma<0$.


We are now ready to investigate the equations of motion generated by
the transformed Hamiltonian (\ref{newHam}). It turns out that in the
vicinity of the singularity they can be solved exactly for many
potentials (see also \cite{ScalarSing} for a general discussion in the
minimally coupled case):
\begin{equation} \label{EoM}
\dot{\tilde{p}} =2 \tilde{p}\f {\tilde{c}}{\gamma}  = \sqrt{\f{2 \kappa}{3}}
\sqrt{\tilde{\pi}^2+2 \tilde{p}^3 \tilde{V}(\tilde{\phi})}
\quad,\quad
\dot{\tilde{\phi}} = \f{\tilde{\pi}}{\tilde{p}}
\quad,\quad
\dot{\tilde{\pi}} = -\tilde{p}^2 \tilde{V},_{\tilde{\phi}}
(\tilde{\phi})
\end{equation}
where the dot denotes a derivative with respect to the conformal time
$\eta$ since we chose $N\propto \sqrt{|\tilde{p}|}$. In the first
equation, we have eliminated the connection $\tilde{c}$, using the
Friedmann constraint $H \approx 0$. We should now specify the form of
the original field potential. The most common choices are the
quadratic and quartic functions $U(\phi)=\f{1}{2}m^2 \phi^2$ or
$U(\phi)=\f{1}{4}\lambda \phi^4$.  The corresponding transformed
potentials then are
\begin{equation} \label{Pot}
\tilde{V}(\tilde{\phi}) =
\f{m^2}{2\sigma}\f{\tanh(\sqrt{\sigma}\tilde{\phi})^2}{\cosh(\sqrt{\sigma}
\tilde{\phi})^2} \quad\mbox{ and }\quad
\tilde{V}(\tilde{\phi}) = \f{\lambda}{4
\sigma^2}\tanh(\sqrt{\sigma}\tilde{\phi})^4\,.
\end{equation}
Note that these expressions are exact and work for any values of the
field, which is one example for simplifications of the canonical
compared to a conformal transformation. In the weak-field limit
($\sigma \tilde{\phi}^2 \ll 1$) one recovers the original
potentials. On the other hand, when approaching the singularity
($\tilde{p} \rightarrow 0$, $\sigma \tilde{\phi}^2
\rightarrow \infty$), the first potential vanishes, whereas the
second one asymptotically tends to a constant, $\frac{1}{4} \lambda
\sigma^{-2}$. In both cases, the potential term in the
$\dot{\tilde{p}}$-equation (\ref{EoM}) can be neglected as
$\tilde{p}\rightarrow 0$, while the righthand side of the
$\dot{\tilde{\pi}}$-equation goes to zero. The latter implies that the
field momentum is a constant, $\tilde{\pi} = \pi_0 = {\rm const}$,
which will thus dominate over $\tilde{p}^3\tilde{V}(\tilde{\phi})$
sufficiently close to $\tilde{p}=0$.  We are, therefore, left with
just two simple equations $\dot{\tilde{p}} = \sqrt{2 \kappa/3} \pi_0$
and $\dot{\tilde{\phi}} = \pi_0/p$.  Assuming that $\t p(\eta=0)=0$,
we get the solutions
\begin{equation} \label{newSols}
\t p(\eta)=\pi_0 \sqrt{\f{2 \kappa}{3}} \eta \propto \eta
\quad,\quad \t \phi (\eta) = \sqrt{\f{3}{2 \kappa}} \ln \eta \propto \ln
\eta\,.
\end{equation}
Remarkably, the solutions do not distinguish between the two
potentials and are recognized as those describing the
massless scalar field. Note, also, that the asymptotic time
dependence of the field does not depend on the initial field
momentum. Converting back to the original variables, we write
\bq \label{sols}
\phi &\equiv&
\f{1}{\sqrt{\sigma}}\sinh(\sqrt{\sigma}\t\phi)=\f{\eta^{\sqrt{\f{3
\sigma}{2 \kappa}}}-\eta^{-\sqrt{\f{3 \sigma}{2
\kappa}}}}{2\sqrt{\sigma}}\propto -\left(\f{1}{\eta}
\right)^{\sqrt{\f{3
\xi}{2}}} \\
p&\equiv&\f{\t p}{f(\phi)} = 4 \pi_0 \sqrt{\f{2 \kappa}{3}}
\f{\eta}{\eta^{\sqrt{\f{3 \xi}{2}}}+\eta^{-\sqrt{\f{3 \xi}{2}}}}
\propto \eta^{1+\sqrt{\f{3 \xi}{2}}} \nonumber
\eq
We conclude the subsection, noting that in the limit of minimal
coupling $\sigma, \xi \rightarrow 0$ (more precisely, $\sigma$ should
go to zero faster than $\phi^{-2}$ to conform with
$\tilde{\phi}\to\infty$) solutions (\ref{sols}) reduce to
(\ref{newSols}). (In the first equation (\ref{sols}), one has to use
the l'Hospital rule when taking $\sigma$ to zero. For doing so, one
must not ignore the first term with a positive power of $\eta$ as we
did in the final step of this equation above at fixed $\sigma$.)

\subsection{Positive coupling constant}

%
%

In this case, the mapping from $\phi$ to $\tilde{\phi}$ is not
one-to-one. Compared to a conformal transformation, one such
singularity at $\Phi_{\rm c}$ is removed which thus turns out to be
spurious and just introduced by the choice of transformation. The
remaining singularity at $\phi_{\rm c}$, however, is different because
the coupling function vanishes at this value, which thus
corresponds to a singular point even in the original system.
Nevertheless, it also appears as a point where the canonical
transformation fails to be one-to-one.

As a consequence, the effective potential $\tilde{V}(\tilde{\phi})$
diverges at this point. There is thus a potential wall which could
simply reflect $\tilde{\phi}$ back when the critical value is
approached and thus restrict the allowed values for $\tilde{\phi}$ to
two disconnected regions. Whether or not this may happen in isotropic
models, an embedding in anisotropic ones shows that $\phi_{\rm c}$
generically corresponds to a shear singularity rather than just a
turning point of $\tilde{\phi}$
\cite{NonMinAniso}.

The irregular potential obtained after the canonical transformation
thus implies a new type of singularity which occurs at a finite value
of $\tilde{\phi}$ where $f(\phi)=0$. As follows from the
considerations in \cite{NonMinAniso}, this happens at a diverging and
possibly non-zero value of $p$. This is a new type of singularity and
presents an interesting test to singularity removal schemes of quantum
gravity: Such schemes are usually based on quantum modifications at
small scales and thus naturally arise around $p=0$. But if variables
are allowed to be of intermediate value at a classical singularity, it
would be difficult for quantum gravity to remove such a singularity
and at the same time preserve classical behavior on large scales.

For a canonical quantization such as loop quantum gravity one uses
canonical variables and it is their appearance at classical
singularities which is relevant. Thus, we have to determine the values
of $\tilde{p}$ and $\phi$ at the new type of singularity when
$\phi=\phi_{\rm c}$. While $\phi$ is finite there, the fact that $p$ does
not diverge together with $f(\phi_{\rm c})=0$ implies that the canonical
variable $\tilde{p}$ must be zero, unlike the geometrical variable
$p$. Thus, singularities always occur in regimes of small $\tilde{p}$
where quantum gravity can dramatically change the behavior and has the
potential to remove singularities.

\section{Loop quantum cosmology}

Loop quantum gravity is based on Ashtekar variables, defined by a
densitized triad and a connection, which allow one to set up a
kinematical quantum framework in a background independent manner.
For minimal coupling, the densitized triad and connection are
conjugate, while for non-minimal coupling the densitized triad has
to be multiplied by the coupling function in order to retain
canonical variables. Upon reducing to isotropic metrics, these
canonical fields directly give the components $\t{p}$ and $\t{c}$
defined before, in addition to the scalar $\phi$ and its momentum
$\pi$. These are the variables that a loop quantization of the
model should be based on since any canonical transformation such
as that to $\t{\phi}$ and $\t{\pi}$ may change properties and may
not be representable by a unitary transformation at the quantum
level.

As in models with minimal coupling \cite{IsoCosmo,Bohr,LivRev}, loop
quantum cosmology in the triad representation is based on wave
functions $\psi(\tilde{\mu},\phi)$ where $\tilde{\mu}$ is a label of
eigenvalues of $\hat{\tilde{p}}$. The classical Friedmann equation for
$\tilde{p}$ is then replaced by a difference equation for the wave
function of the type
\begin{equation}\label{DiffEq}
 C_1(\tilde{\mu})\psi(\tilde{\mu}+1,\phi)-2
C_0(\tilde{\mu})\psi(\tilde{\mu},\phi)+
C_{-1}(\tilde{\mu})\psi(\tilde{\mu}-1,\phi)\propto \hat{H}_{\rm
matter}(\tilde{\mu}) \psi(\tilde{\mu},\phi)
\end{equation}
with the matter Hamiltonian operator $\hat{H}_{\rm
matter}(\tilde{\mu})$ which for standard matter is diagonal on states
$|\tilde{\mu}\rangle$. In the limit of large scale factors $\t \mu \gg
1$, this difference equation yields the Wheeler-DeWitt differential
equation while at small scales the discreteness of quantum geometry is
essential. The criterion for non-singular behavior (see also
\cite{DegFull} for details) then consists in unique extendability of
the wave function even across classical singularities for specific
values of $\tilde{\mu}$. In the standard case, singularities are
reached for $\tilde{\mu}=0$, where the kinematical structure provides
a new branch since there are two sides to the classical singularity
corresponding to opposite orientations of the triad. While they cannot
be connected classically, it is possible to extend the wave function
uniquely.  To see this, one has to use a more detailed form of
coefficients of the difference equation and discuss when they become
zero. It turns out that, although leading coefficients $C_{\pm 1}$ can
become zero, this does not spoil the recurrence scheme of the
difference equation. The wave function is then uniquely extended and
quantum gravity can tell us about the other side beyond singularities
\cite{Sing}.

For this argument it is important to provide a candidate for the other
side, such as $\tilde{\mu}<0$, and then show that dynamics extends the
wave function uniquely. Thus, this scheme only works when classical
singularities are indeed located at values of vanishing $\tilde{\mu}$
(or degenerate triads) since there would be no new region
otherwise. While this is automatically the case in isotropic models
with standard types of matter fields, it is non-trivially realized in
anisotropic models \cite{HomCosmo} or even spherical symmetry
\cite{SphSymmSing}. In anisotropic models, for instance, the Kasner
behavior where one metric component always diverges at a singularity
suggests a general behavior where not all metric components vanish at
singularities. However, it turns out that densitized triad components
which underlie the loop quantization always vanish at anisotropic
classical singularities. Thus, the same argument to remove classical
singularities applies. This is a general scheme, but non-minimally
coupled models were not included so far. In fact, a second crucial
ingredient is the assumption of a matter Hamiltonian operator diagonal
on triad eigenstates. Classically, this corresponds to a matter
Hamiltonian which depends only on spatial geometry but not on
curvature. If there are curvature terms and $\hat{H}_{\rm matter}$ is
not diagonal in geometric variables, the right hand side of
(\ref{DiffEq}) could also contribute terms proportional to
$\psi(\tilde{\mu}\pm 1,\phi)$ which would change the recurrence and
introduce new places where the leading coefficients vanish.  The
recurrence would then have to be re-analyzed and singularity removal
would be less general since the coefficients would depend on the
matter field through the Hamiltonian, and not just on $\tilde{\mu}$.

We can now test this scheme with what we have learned about
non-minimally coupled isotropic models.  First, there may be
singularities where even the geometric triad variable $p$ may not
vanish. However, it turned out that the canonical variable $\tilde{p}$
does vanish. Since the difference equation is based on the
quantization of this variable, non-minimally coupled models are
included in the non-singularity scheme of loop quantum cosmology. As
the discussion here illustrates, this happens in a way depending on
dynamical details of the models.

At first sight, one could also expect that curvature
couplings of a non-minimal model lead to non-diagonal matter
Hamiltonians since there are curvature components in the matter
terms. Here, the previous expressions show that the matter Hamiltonian
(\ref{HamCanon}) in canonical variables does depend only on
$\tilde{p}$ in addition to the matter fields, but not on the
connection component $\tilde{c}$ which determines curvature. Thus, the
matter Hamiltonian even with non-minimal coupling is diagonal in the
canonical triad representation and the general singularity removal
scheme applies.  This remains true for positive spatial curvature not
spelled out explicitly here, since we would only have to replace
$\tilde{c}^2$ by $\tilde{c}^2+1$ \cite{Closed}. Thus, non-minimally
coupled isotropic models are non-singular in loop quantum cosmology in
the same manner as minimally coupled ones.

\section{Conclusions}

Applying a canonical transformation to a non-minimally coupled
isotropic cosmological model we have shown that for classically
allowed coupling functions singularities occur as in the minimally
coupled case. For quadratic and quartic scalar potentials, moreover,
the asymptotic approach to a singularity is as in the massless, free
minimal case.

This has direct implications for the question of singularity removal
in quantum gravity. Loop quantum cosmology \cite{LivRev} has provided
a scheme of singularity removal which has been shown previously to
apply to isotropic \cite{Sing}, diagonal class A Bianchi models
\cite{HomCosmo,Spin}, and spherically symmetric and polarized
cylindrical wave models \cite{SphSymmSing,SphSymmHam}. In all these
cases, it was shown that quantum dynamics provides a well-posed
recurrence scheme for wave functions across classical
singularities. It was therefore important to know at which geometrical
configurations a classical singularity occurs, such as at vanishing
scale factor $p=0$ for isotropic models coupled to a scalar.  All
these arguments were independent of the form of the matter Hamiltonian
and thus present a quantum geometry effect, provided that the
classical matter Hamiltonian did not depend on curvature.

For the purpose of quantum gravity, curvature couplings fall into
two different classes, depending on whether they require a
non-zero Planck length or not.  Most curvature coupling terms in
effective matter actions require dimensionful coefficients with a
new length scale to compensate the dimensions of curvature
factors. Such terms cannot arise in the classical Hamiltonian used
to set up a quantization. Such actions with curvature coupling
terms will thus not be loop-quantized; they would rather follow
from a quantum theory of gravity such as loop quantum gravity in
an effective description, see e.g.\ \cite{Karpacz}. The curvature
coupling of a non-minimal scalar, on the other hand, is allowed
classically provided that the coupling function is quadratic in
$\phi$, but it was not covered by previous discussions of
singularities in the context of loop quantum gravity.

With the results of the present paper, we can fill this gap. There are
two potential problems: the Hamiltonian could be curvature dependent,
and different types of singularities could occur.  For the classically
allowed form of coupling function which is quadratic, we have seen
that classical singularities always occur at vanishing
$\tilde{p}=f(\phi)p$ which is one of the canonical variables, in the
same form as in minimal models. Moreover, the form of quantum dynamics
near classical singularities is the same whether a scalar is coupled
minimally or non-minimally. The argument of singularity removal is
thus unchanged.

The discussion of this paper shows that this conclusion is
realized quite non-trivially: The classically allowed case of a
quadratic coupling function is just the limiting case where an
infinite scalar $\phi$ is mapped to an infinite $\t\phi$ by the
canonical transformation. For coupling functions of higher degree,
the possibility of singularities at finite values of $\t\phi$ and
non-zero $\t p$ cannot easily be ruled out. In such a case, the
usual singularity removal mechanism and the resulting non-singular
picture of quantum geometry would have been very different, if
applicable at all. Since higher degrees than two in the coupling
function can only arise effectively, but not in a classical action
yet to be quantized, the singularity structure relevant for
isotropic loop quantum cosmology does not change for non-minimal
coupling. As with calculations of black hole entropy
\cite{NonminScalar} we have seen that, although the non-minimal
situation appears initially crucially different from the minimal
one, the methods of loop quantum gravity are general enough to
encompass automatically also non-minimal situations in quite
non-trivial ways.

\section*{Acknowledgements}

We thank Alexey Toporensky for discussions and for pointing out
Refs.~\cite{ScalarSing,ScalarSingWall} to us.

\end{document}